\documentclass[lettersize,journal]{IEEEtran}
\usepackage{amsmath,amsfonts,amssymb} 
\usepackage{algorithm}
\usepackage{algpseudocode}
\usepackage{array}
\usepackage[caption=false,font=footnotesize,labelfont=rm,textfont=rm]{subfig}
\usepackage{textcomp}
\usepackage{stfloats}
\usepackage{url}
\usepackage{verbatim}
\usepackage{graphicx}
\usepackage{cite}
\usepackage{hyperref}

\usepackage{xcolor}
\usepackage{color}
\ifodd 1

  \newcommand{\com}[1]{\textbf{\color{red}(COMMENT: #1)}} 
\else

  \newcommand{\com}[1]{}
  
  \fi
\ifodd 0
  
\else
  
  \fi

\begin{document}

\title{
Explicit Semantic-Base-Empowered Communications for 6G Mobile Networks
}

\author{Fengyu Wang, Yuan Zheng, Wenjun Xu, Junxiao Liang, Ping Zhang, and Zhu Han
\thanks{This work was supported in part by the National Natural Science Foundation of China (62293485 and 62301069). Corresponding author: Wenjun Xu (wjxu@bupt.edu.cn).

Fengyu Wang is with the School of Artificial Intelligence, Beijing University of Posts and Telecommunications, Beijing 100876, China.
Yuan Zheng, Wenjun Xu, Junxiao Liang, and Ping Zhang are with the State Key Laboratory of Networking and Switching Technology, Beijing University of Posts and Telecommunications, Beijing 100876, China. 
Zhu Han is with the Department of Electrical and Computer Engineering, University of Houston, Houston, TX 77004, USA.
}}

\maketitle

\begin{abstract}
Increasing demands for massive data transmission pose significant challenges to communication systems. Compared with traditional communication systems that focus on the accurate reconstruction of bit sequences, semantic communications (SemComs), which aim to deliver information connotation, are regarded as a key technology for sixth-generation (6G) mobile networks. Most current SemComs utilize an end-to-end (E2E) trained neural network (NN) for semantic extraction and interpretation, which lacks interpretability for further optimization. Moreover, NN-based SemComs assume that the application and physical layers of the protocol stack can be jointly trained, which is incompatible with current digital communication systems. To overcome those drawbacks, we propose a SemCom system that employs explicit semantic bases (Sebs) as the basic units to represent semantic connotations. First, a mathematical model of Sebs is proposed to build an explicit knowledge base (KB). Then, the Seb-based SemCom architecture is proposed, including both a communication mode and a KB update mode to enable the evolution of communication systems. Sem-codec and channel codec modules are designed specifically, with the assistance of an explicit KB for the efficient and robust transmission of semantics. Moreover, unequal error protection (UEP) is strategically implemented, considering communication intent and the importance of Sebs, thereby ensuring the reliability of critical semantics. In addition, a Seb-based SemCom protocol stack that is compatible with the fifth-generation (5G) protocol stack is proposed. To assess the effectiveness and compatibility of the proposed Seb-based SemComs, a case study focusing on an image-transmission task is conducted. The simulations show that our Seb-based SemComs outperform state-of-the-art works in learned perceptual image patch similarity (LPIPS) by over 20\% under varying communication intents and exhibit robustness under fluctuating channel conditions, highlighting the advantages of the interpretability and flexibility afforded by explicit Sebs. 
\end{abstract}


\section{Introduction}

\IEEEPARstart{O}{ver} the past few decades, the rapid development of mobile communications has profoundly revolutionized human society, shifting the world from the Internet of Things (IoT) to the Internet of Everything (IoE). A variety of innovative services, such as extended reality (XR) and digital twins (DT), have emerged, with escalating requirements for data throughput and ubiquitous intelligence, catalyzing the development of sixth-generation (6G) mobile networks.

Current state-of-the-art communication frameworks are mainly designed based on Shannon’s classical information theory (CIT) \cite{shannon1948mathematical}, which characterizes information as random variables and assumes that semantics are irrelevant to communications, for the sake of the generality of the theory \cite{semcom_survey_2023}. However, 6G envisions the integration of ubiquitous artificial intelligence (AI) terminals to build a network that incorporates not only communications but also perception and intelligence in order to achieve ubiquitous connectivity and intelligence. Consequently, CIT-based communication systems are struggling to meet the grand vision of 6G, necessitating a fundamental shift in the communication paradigm \cite{paradigm-shift}. Thanks to recent success in AI, user equipment (UE) such as IoT sensors and smartphones can understand the meaning of raw data and use it to perform intelligent tasks \cite{semcom-edge}. As a result, semantic communications (SemComs), which perform in an “understanding-before-transmission” manner \cite{understand-before-talk}, have been rapidly developing in recent years and are considered a key technology in achieving the vision of 6G \cite{zhang_survey}.

\begin{figure}[!t]
\centering
\includegraphics[width=0.49\textwidth]{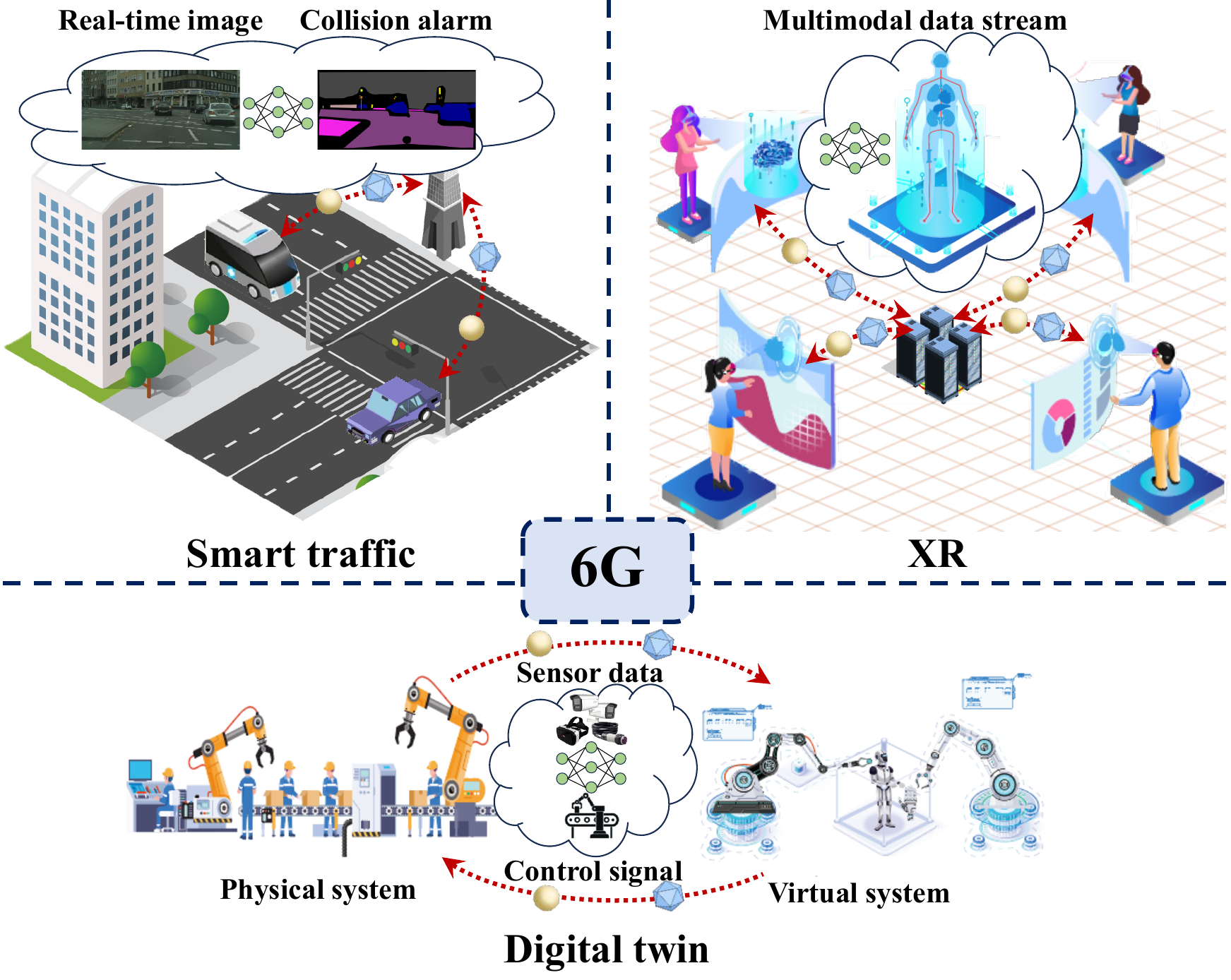}
\caption{SemCom-empowered services for typical 6G applications. }
\label{fig:6gsemcom}
\end{figure}

SemComs are capable of adaptively constructing scenario-specific services by processing the essential connotations of messages according to the intent of intelligent tasks (Fig. \ref{fig:6gsemcom}), thereby increasing the intelligence and efficiency of the system. More specifically, the meaning of messages is extracted and sent in a form that is understandable for both source and destination, with a knowledge base (KB) being introduced in SemComs to ensure transceivers share the same background knowledge. A prevalent approach employed in recent works involves the establishment of an implicit KB based on an end-to-end (E2E) trained neural network (NN) for SemComs \cite{bourtsoulatze2019deep, zhijin_text, cross-modal}, also known as deep joint source and channel coding (DeepJSCC). To ensure transceivers can understand each other, NNs at both the transmitter and receiver are jointly trained in semantic encoding/decoding (codec), where the parameters of the NNs serve as the implicit KB of SemComs.

Although techniques such as DeepJSCC have demonstrated the superiority of SemComs in terms of communication efficiency, the implicit KB-based architecture presents several challenges for practical implementation. A primary issue is that these systems are trained under the premise that the environment (i.e., the statistics of messages and channels) is static \cite{nine}. If channel conditions or the knowledge within messages deviate from the scope of the training data, the entire system needs to be retrained, resulting in considerable additional overheads \cite{Ren_knowledge, Yi_Deep}. To overcome this issue, a task-unaware SemCom system for image transmission is proposed in Ref. \cite{zhang2023unaware}, where transfer learning is employed to evolve the KB, adapting the system under dynamic data settings. Update mechanisms for the encoder–decoder NN and the KB module are proposed in Ref. \cite{luo_semantic}, in which KB is modeled as an explicit module independent of the encoder–decoder structure. However, both studies \cite{zhang2023unaware, luo_semantic}, indirectly treat the training dataset as the KB, making the system’s utilization of knowledge indirect and necessitating an additional fine-tuning process. Moreover, the aforementioned approaches struggle with issues of interpretability, as the lack of interpretability makes it difficult to measure the effectiveness of systems and make relevant updates. To theoretically clarify the interpretability of SemComs, a mathematical theory of SemComs is proposed in Ref. \cite{niu2024mathematical}, which models SemComs from the perspective of synonymous mapping. However, the question of how to implement SemComs following the proposed explicit and controllable synonymous mapping has not yet been comprehensively studied.

In addition, thanks to transceivers’ ability to understand the meaning of the source message in SemComs, implementing unequal error protection (UEP) based on varying levels of message importance for a specific task has become an effective way to increase the efficiency of SemCom \cite{dai_communication, fu_scalable}. Semantic importance can be detected by entropy modeling \cite{balle_nonlinear}, saliency detection \cite{judd_learning}, gradient-weighted class activation mapping (Grad-CAM) \cite{selvaraju2017grad}, and so forth. However, most studies apply UEP based on an implicit KB; as a result, their importance modeling is mainly limited to the individual messages themselves. Additionally, most current joint source and channel coding (JSCC)-based SemComs assume that the communication systems support full-resolution constellation points \cite{zhang2024analog} and therefore directly map semantics to continuous complex constellations. As a result, it is difficult to make these JSCC-based SemComs compatible with current digital communication systems. To achieve compatibility, DeepJSCC-Q-type techniques \cite{zhang2024analog, tung2022deepjscc} set constellation points as finite and use an NN to learn constellation mapping. However, the modulation scheme cannot change once the model is trained, resulting in a lack of flexibility in practical use. Moreover, SemComs based on DeepJSCC assume that the application layer and physical layer can be jointly trained, ignoring the constraints proposed by current protocol stacks. Therefore, the question of how to design compatible SemCom systems with a higher degree of interpretability and flexibility is an open yet important topic for 6G.

To this end, in this article, we propose a new SemCom architecture based on explicit semantic bases (Sebs), where the generation of Sebs serves as the direct realization of synonymous mapping \cite{seb-zhang}. Moreover, two levels of UEP—that is, Seb-wise and message-wise UEP—are proposed to improve the transmission efficiency of SemComs. Protocol stacks compatible with CIT-based systems are further proposed. The contributions of this work are summarized as follows:
\begin{itemize}
    \item A mathematical model of Sebs is proposed, laying the foundation for building an explicit KB for SemComs. Based on this, a SemCom architecture with an explicit KB is proposed, consisting of both a communication mode and an update mode, enabling the intelligent evolution of the system. 
    \item Criteria for designing a semantic encoder/decoder (Sem-codec) and channel codec are illustrated, where the communication intent is incorporated through the Seb importance level and messages to provide UEP for critical semantics. 
    \item To facilitate the practical implementation of SemComs, a Seb-based SemCom protocol stack consistent with the current 3rd Generation Partnership Project (3GPP) protocol stack is proposed, bridging the gap between theory and application with respect to SemComs for 6G mobile networks.
\end{itemize}

The remainder of the paper is organized as follows: Section \ref{sec-ii} introduces the architecture of the Seb-based SemComs, including Sebs, the KB, the Sem-codec, and the channel codec. Next, the protocol of Seb-based SemComs is further discussed in Section \ref{sec-iii}, followed by a case study to demonstrate the effectiveness of the proposed Seb-based SemComs in Section \ref{sec-iv}. The work is concluded in Section \ref{sec-v}.

\section{Architecture of SemComs with explicit Sebs}\label{sec-ii}

\begin{figure*}[!t]
    \centering
    \includegraphics[width=0.9\textwidth]{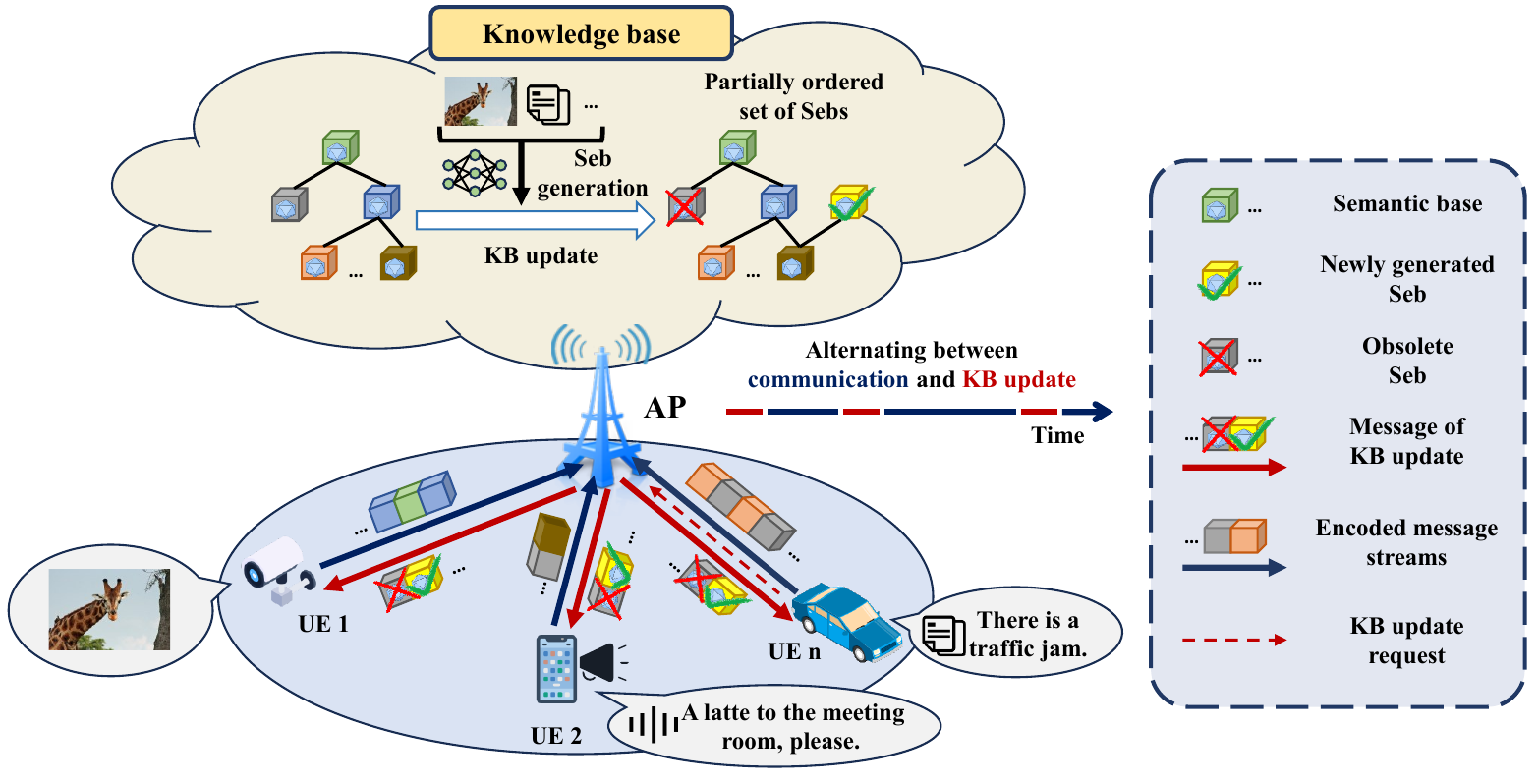}
    \caption{Architecture of the Seb-based SemComs. The system performs in two modes: the communication mode and the KB update mode. During communications, messages are interpreted based on a synchronized KB. A KB update is triggered through requests from the UEs to evolve the system. AP: access point. }
    \label{fig:scenarios}
\end{figure*}

In this section, we provide an overview of the proposed Seb-based SemCom architecture. As shown in Fig. \ref{fig:scenarios}, the architecture involves two modes—namely, the communication mode and the KB update mode—enabling the proposed SemCom to provide intelligent services while being capable of self-evolution and optimization. During communications, the KB of the access point (AP) and that of the UEs are synchronized, facilitating the extraction and interpretation of message semantics. Notably, the underlying intent of communications may change over time, making a KB obsolete and resulting in inefficient semantic representation (SR). Consequently, updating is necessary to enable the evolution of the network. The KB update mode is triggered by requests from the UEs, during which the AP generates and transmits a KB update message by analyzing the recently sent messages. To enable flexible SR and efficient KB updates, the KB is composed of explicit Sebs, represented by the diamond-marked cubes in Fig. \ref{fig:scenarios}. In the remainder of this section, the mathematical model of Sebs will be introduced in Section \ref{sec-ii-seb}, based on which the criteria for designing the KB, Sem-codec, and channel codec will be further illustrated in Sections \ref{sec-ii-kb}, \ref{sec-ii-sem}, and \ref{sec-ii-uep}, respectively.

\subsection{Semantic base}
\label{sec-ii-seb}

The introduction of Sebs in the KB grants SemCom systems great interpretability and flexibility compared with implicit KB-based approaches. As the basic unit for expressing semantic information, Sebs serve as an analogous concept to bits in CIT \cite{seb-zhang}. Nonetheless, the granularity of Sebs can be adjusted to align with the intent of communication for intelligent services, thereby meeting diverse representational demands.

\begin{figure}[!t]
    \centering
    \includegraphics[width=0.48\textwidth]{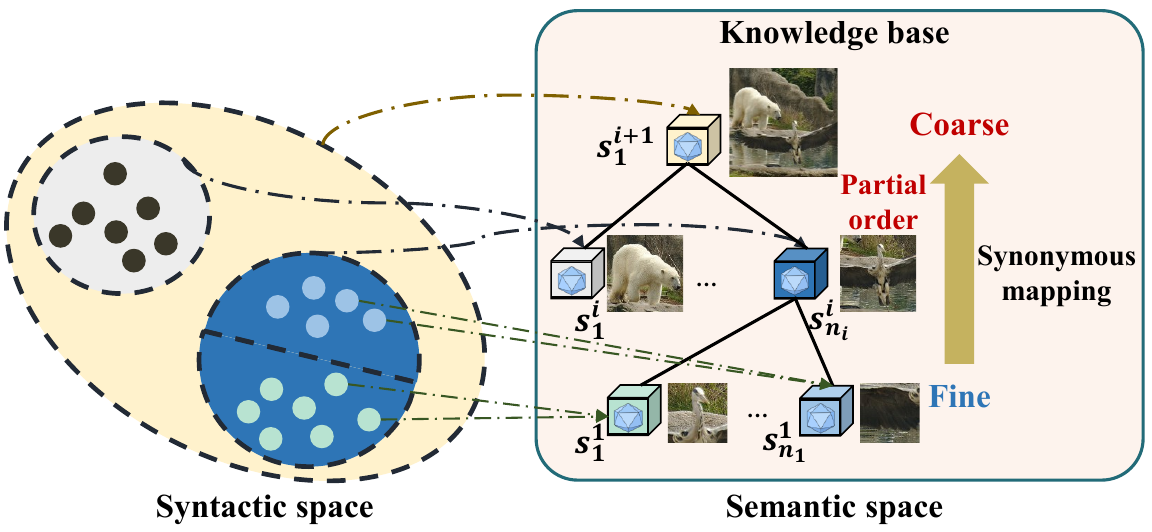}
    \caption{Illustration of a partially ordered set of Sebs, where the semantics of the original message at various granularities are depicted by Sebs at different levels. 
    $\boldsymbol{s}^i_{n_i}$: the $n_i$th Seb at the $i$th level; 
    $n_i$: the total number of Sebs in $\mathcal{S}^i$;
    $i\in\{1,2,...\}$: the level index;
    $S^i$: the generated set of Sebs at the $i$th level.}
    \label{fig:syn_map}
\end{figure}

We regard the generation of Sebs as a concrete form of synonymous mapping \cite{niu2024mathematical}, where messages sharing the same meaning in multiple manifestations can be represented by the same Seb. In other words, the relationship between syntactic information and its corresponding semantics is typically many-to-one mapping \cite{niu2024mathematical}, as shown in Fig. \ref{fig:syn_map}. In practice, clustering algorithms could be employed to realize the aforementioned process of synonymous mapping. By clustering the raw semantic features of message segments with similar semantics, the corresponding cluster centers can effectively summarize the emerged ``synonyms''. Given a set of accessible messages $\mathcal{X} = \{ \tilde{\boldsymbol{x}}_1,\tilde{\boldsymbol{x}}_2,...,\tilde{\boldsymbol{x}}_l,...\}$, where $\tilde{\boldsymbol{x}}_l$ denotes the $l$th message and $l\in\{ 1,2,...\}$ is the message index, this process can be mathematically expressed as follows:
\begin{equation}
    \mathcal{S}^1 = f_{\mathrm{cluster}}({f_{\mathrm{SE}}}(\mathcal{X})) = f_{\mathrm{cluster}}(\mathcal{F}),
    \label{eq:clus}
\end{equation}
where $f_\mathrm{SE}(\cdot)$ denotes the semantic extraction module, which can be implemented using a pretrained NN to output a collection of raw semantic features $\mathcal{F}$. $f_\mathrm{cluster(\cdot)}$ denotes the clustering algorithm employed (e.g., \textit{K}-means). The variable $\mathcal{S}^1 = \{ \boldsymbol{s}^1_1,\boldsymbol{s}^1_2,...\boldsymbol{s}^1_k,...,\boldsymbol{s}^1_{n_1}\}$ represents the generated set of Sebs at the first level, where $n_1$ represents the total number of Sebs in $\mathcal{S}^1$, and $\boldsymbol{s}^1_k$ denotes the $k$th Seb in $\mathcal{S}^1$ ($k\in\{1,2,...,n_1\}$).

A synonymous mapping correlation also exists between fine- and coarse-grained semantics. For instance, in scenarios requiring coarser Seb granularity, such as higher-level intelligent tasks including classification or object detection, multiple fine-grained Sebs can be further mapped into a single coarse-grained Seb based on synonymous mapping, thereby leading to a more abstract representation of information for transmissions (shown in Fig. \ref{fig:syn_map} as Sebs in higher levels). We can iteratively cluster the Sebs obtained from Eq. (\ref{eq:clus}) to derive higher-level Sebs; that is,
\begin{equation}
    \mathcal{S}^{i+1} = f_{\mathrm{cluster}}(\mathcal{S}^i).
    \label{eq:clus_higher}
\end{equation}
Here, $\mathcal{S}^i$ with superscripts $i\in\{1,2,...\}$ represents the set of Sebs at the $i$th level. In other words, Sebs in different granularities exhibit a hierarchical structure. Accordingly, we model the set of Sebs as a partially ordered set $\langle \mathcal{S},\preccurlyeq \rangle$, where Sebs are elements in the set of Sebs $\mathcal{S}$, and $\preccurlyeq$ denotes the synonymous mapping relationship, indicating the process of mapping fine-grained Sebs into coarse-grained Sebs, as shown in the Hasse diagram in Fig. \ref{fig:syn_map}.

\begin{figure*}[!t]
    \centering
    \includegraphics[width=0.9\textwidth]{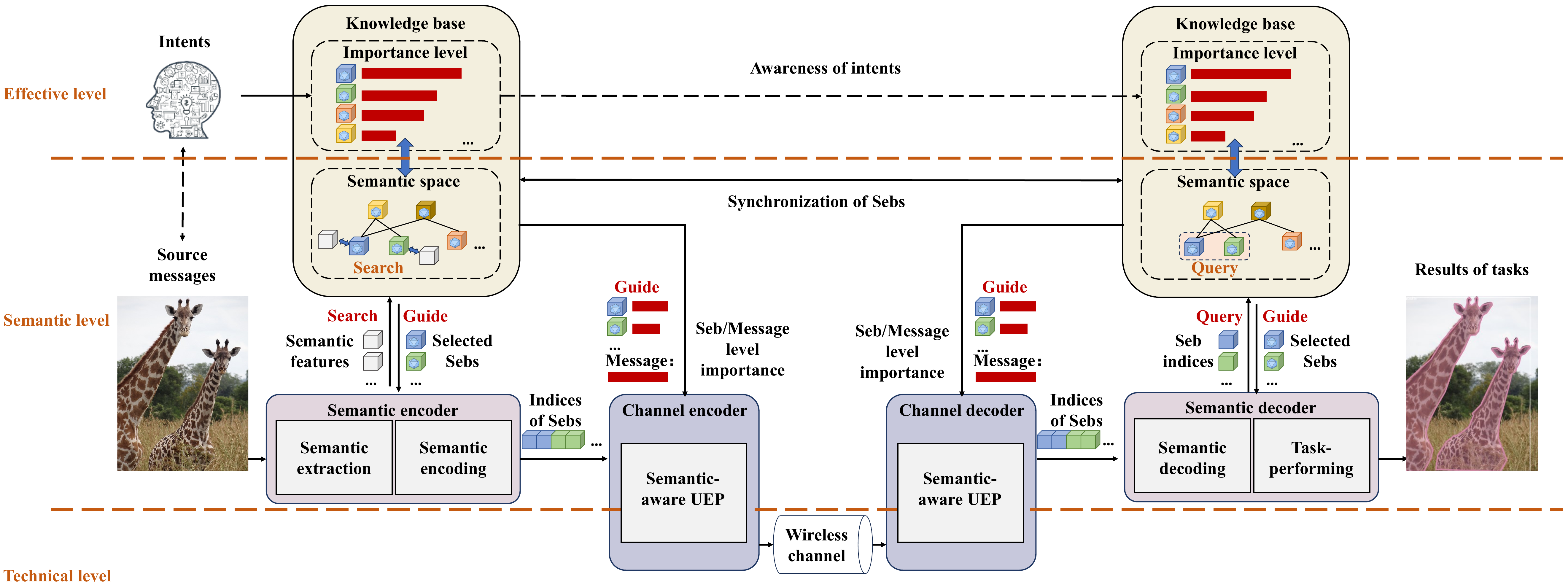}
    \caption{Architecture of the proposed SemCom system. Sebs are regarded as the basic unit for expressing semantic information and are synchronized in the KB to support the implementation of the Sem-codec and channel codec. }
    \label{fig:framework}
\end{figure*}

Fig. \ref{fig:syn_map} gives an exemplary process of Seb generation for an image, where syntactic information that records pixel-level details corresponds to the finest granularity points in the feature space, shown as dots in the left part. Fine-grained Sebs aggregate groups of pixel-level details sharing the same meaning, corresponding to the lowest row of Sebs in the KB. Coarse-grained Sebs, shown in the first and the second rows in the KB, aggregate sets of similar fine-grained Sebs, forming more abstract semantic concepts corresponding to messages that share the same meaning in a broader perspective (i.e., the dots in the dashed circles in the left part of the figure). Overall, by leveraging the many-to-one correlation between the source messages and their corresponding semantics across different granularities, Sebs can condense multiple manifestations of the same meaning into a unified SR, thereby improving efficiency.

The encoding process utilizing Sebs will be described further in Section \ref{sec-ii-sem}. In practical systems, Seb granularity is determined by both user intent and the available transmission resources. For certain intelligent tasks, we can assign varying levels of importance to different fractions of the transmitted message, considering their differentiated representational needs. More specifically, coarse-grained Sebs excel at capturing abstract semantics, which are suitable for resource-constrained scenarios to support the transmission of essential task-related semantics. In comparison, fine-grained Sebs, which incur higher communication overheads, are more accurate in SR and are ideal for conveying detailed semantics according to the communication intent. The combination of Sebs with different granularities offers a flexible and efficient methodology for the execution of intelligent tasks. Moreover, it should be noted that Sebs are not limited to representing messages within the same modality. By customizing feature-extraction methods for different modalities, a shared feature space can be built that jointly aligns semantics from different modalities. Consequently, Sebs can be generated following the same procedure shown in Fig. \ref{fig:syn_map}, providing a shared SR for data in various modalities. In other words, Sebs can be used to facilitate the alignment of multimodal data, which would be significant for the realization of ubiquitous connectivity and intelligence for novel applications in 6G networks.

\subsection{Knowledge base}
\label{sec-ii-kb}
The KB is a distinctive component of SemComs that has no counterpart in CIT-based networks. By synchronizing knowledge across transceivers, the KB enables messages to be interpreted with Sebs that serve as shared prior knowledge, thus ensuring the effectiveness of SemComs.

It should be noted that different fractions of the transmitted message may have varying levels of importance, and the importance of Sebs should differ accordingly. For example, in a scenario where a set of images are transmitted for object recognition, Sebs that mainly represent the images’ foreground should be more important than those mainly representing the background. Moreover, communication intents may often change dynamically in practical scenarios, leading to a mismatch between statically generated Sebs and dynamic intents and decreasing the accuracy and efficiency of SR. Therefore, dynamic updates of Sebs are essential and should further consider Seb importance levels, which are shown as the red bars beside the Sebs in Fig. \ref{fig:framework}.

To be specific, the importance levels of Sebs are first initialized based on the importance of the Sebs’ corresponding raw semantic features or lower-level Sebs; that is,
\begin{equation}
  I_k^1 = \frac{1}{|\mathcal{F}_k|}\sum\limits_{\boldsymbol{f} \in \mathcal{F}_k} I_{\boldsymbol{f}},
  \label{imp}
\end{equation}
where $I_k^1$ denotes the importance of $\boldsymbol{s}_k^1$. $\mathcal{F}_k \subseteq \mathcal{F}$ is the set of features clustered into $\boldsymbol{s}_k^1$, $|\cdot|$ denotes the cardinality of the set, and $I_{\boldsymbol{f}}$ is the importance of the raw semantic feature $\boldsymbol{f}$ that can be obtained through methods such as Dino \cite{oquab2024dinov2} or Grad-CAM \cite{selvaraju2017grad}. Additionally, we have
\begin{equation}
  I_k^{i+1} = \frac{1}{|\mathcal{S}_k^i|}\sum\limits_{\boldsymbol{s}_m^i\in \mathcal{S}_k^i} I_m^i,
  \label{imp_higher}
\end{equation}
for higher-level Sebs with $i\in\{2,3,...\}$. Similarly, $\mathcal{S}^i_k \subseteq \mathcal{S}^i$ is the set of Sebs clustered into $\boldsymbol{s}_k^{i+1}$, $I_m^i$ denotes the importance of the $m$th Seb in the $i$th level ($m\in\{1,2,...,n_i\}$), which is further clustered into $\boldsymbol{s}_k^{i+1}$. $I_k^i$ is the importance of the $k$th Seb in the $i$th level.

To ensure the efficiency of the proposed SemCom system, a KB update is triggered when the UE detects inaccuracy in the SR, as indicated by the dashed line in Fig. \ref{fig:scenarios}. In implementation, the accuracy of the SR can be measured by the distortion  between the raw semantic features of a message to be transmitted $\boldsymbol{x}$ and the Sebs closest to them—that is, the Euclidean distance:
\begin{equation}
D({f_{{\mathrm{SE}}}}(\boldsymbol{x}),\mathcal{S}) = {\left\| {{f_{{\mathrm{SE}}}}(\boldsymbol{x})-{f_{{\mathrm{quant}}}}({f_{{\rm{SE}}}}(\boldsymbol{x}),\mathcal S}) \right\|_2},
\label{eq:acc}
\end{equation}
where $f_\mathrm{quant}(\cdot)$ is the semantic encoding module that identifies the closest Sebs in a predefined level based on the importance of the raw semantic features. When the accuracy falls below the predefined threshold, the UE will regard the current SR as inaccurate. The update process includes the following strategies to capture emerging topics:
\begin{itemize}
    \item \textbf{Add new Sebs}. New candidate Sebs are generated from recent messages, along with their corresponding levels of importance. These candidates are integrated into the KB if they exhibit significant differences from the existing Sebs (e.g., may also be measured by the Euclidean distance).
    \item \textbf{Adjust Seb importance.} The levels of importance of existing Sebs are further adjusted according to their age of information (AoI) \cite{semcom-edge}, which underscores the importance of freshness. Specifically, the level of importance of a Seb decreases over time but will be refreshed upon usage; that is,
    \begin{equation}
        I_k^i(t) = I_k^i \cdot \Delta (t - {t_0}),\quad t \ge {t_0},
        \label{eq:adj_imp}
    \end{equation}
    where $\Delta(t-t_0)$ is a monotonically decreasing function satisfying $\Delta(0)=1$ and $\Delta(+\infty)=0$. Here, $t_0$ denotes the most recent time at which Seb $\boldsymbol{s}_k^i$ was used, and $t$ is the current time stamp. 
    \item \textbf{Remove obsolete Sebs.} Sebs with levels of importance falling below a predefined threshold are considered obsolete and are removed from the KB, thereby reducing KB redundancy and improving encoding efficiency.
\end{itemize}

The KB update messages are synchronized from the AP to the UEs, as shown in Fig. \ref{fig:scenarios}.

\subsection{Semantic encoder/decoder}\label{sec-ii-sem}

The Sem-codec is deployed to extract and interpret the source message utilizing Sebs, which can be realized by an NN. The semantic encoder mainly comprises a semantic extraction module $f_\mathrm{SE}(\cdot)$ and a semantic encoding module $f_\mathrm{quant}(\cdot)$, as shown in Fig. \ref{fig:framework}. The source message is first mapped into the semantic space through semantic extraction, yielding the corresponding hierarchical semantic features. Then, the semantic features are quantized by corresponding Sebs within the KB. Based on their levels of importance, these features are quantized using Sebs with different granularities. To be specific, more important features are quantized using finer Sebs, while less important features are quantized using coarser Sebs. The semantic encoder $f_\mathrm{sem}^\mathrm{en}(\cdot)$ can be expressed as follows: 
\begin{equation}
    \boldsymbol{s}^{\mathrm{idx}} = f_{\mathrm{sem}}^{\mathrm{en}}(\boldsymbol{x},\mathcal{S}) = {f_{\mathrm{quant}}}({f_{\mathrm{SE}}}(\boldsymbol{x}),\mathcal{S}),
    \label{eq:semen}
\end{equation}
where the output $\boldsymbol{s}^\mathrm{idx}$ records the indices of the selected Sebs. The semantic decoder consists of a semantic decoding module $f_\mathrm{query}(\cdot)$ and a downstream task-performing module $f_\mathrm{TP}(\cdot)$. The received $\hat{\boldsymbol{s}}^\mathrm{idx}$ are first interpreted by $f_\mathrm{query}(\cdot)$ with the assistance of the KB, which retrieves the semantic features according to the Seb indices. Then, the semantic features are directly used by the task-performing module to execute downstream tasks; that is,
\begin{equation}
    \boldsymbol{z} = f_{{\mathrm{sem}}}^{{\mathrm{de}}}(\hat{\boldsymbol{s}}^{\mathrm{idx}},\mathcal{S}) = {f_{\mathrm{TP}}}({f_{\mathrm{query}}}(\hat{\boldsymbol{s}}^\mathrm{idx},\mathcal{S})),
\end{equation}
where $\boldsymbol{z}$ is the task-performing result and $f_{{\mathrm{sem}}}^{{\mathrm{de}}}(\cdot)$ represents the semantic decoder.

Since the accuracy and efficiency of SR vary with the granularity of Sebs, by considering the specific requirements of the communication intent, a balance can be achieved through the comprehensive application of Sebs with different granularities. Fig. \ref{fig:framework} illustrates a SemCom system in which images are transmitted for classification. Fine-grained Sebs can be adopted to represent the key features necessary for distinguishing between categories, such as the facial and body patterns of a giraffe, which ensures overall accuracy. Meanwhile, coarse-grained Sebs can be used for less critical features, such as the trees and the sky in the background, thereby increasing the efficiency of the communication process. 

\subsection{Channel encoder/decoder}\label{sec-ii-uep}

It should be noted that, while Sem-codec determines the encoding efficiency of the source, the transmission accuracy needs to be further considered in practical applications. In general, the transmission process of the encoded symbol$\boldsymbol{s}^\mathrm{idx}$ through the channel codec can be expressed as follows:
\begin{equation}
\begin{aligned}
\boldsymbol{y}& = f_{\mathrm{chan}}^{\mathrm{en}}({\boldsymbol{s}^{\mathrm{idx}}},\mathcal{I}),\\
\hat{\boldsymbol{s}}^{\mathrm{idx}}& = f_{\mathrm{chan}}^{\mathrm{de}}(\hat{\boldsymbol{y}},\mathcal{I}),
\label{chanen}
\end{aligned}
\end{equation}
where $f^\mathrm{en}_\mathrm{chan}(\cdot)$ and $f^\mathrm{en}_\mathrm{chan}(\cdot)$ denote the channel encoding and decoding modules, respectively; $\mathcal{I} = \{ I_1^1(t),I_2^1(t),...,I_n^i(t),I_{n + 1}^i(t),...\} $ represents the set of importance levels of Sebs; $\boldsymbol{y}$ denotes the channel encoding result of $\boldsymbol{s}^\mathrm{idx}$; and $\hat{\boldsymbol{y}}$ is the impaired version of $\boldsymbol{y}$ after transmission over the physical channel. Similar to the design in the Sem-codec, semantic information with different levels of importance have differing requirements for accuracy: Errors in important semantics must be minimized, whereas less-important semantics have a certain level of error tolerance in transmission. More specifically, elements in $\boldsymbol{s}^\mathrm{idx}$ that are assigned different importance levels should be encoded with different channel-coding strategies, where the UEP mechanism can be naturally introduced. For Seb-based SemCom systems, UEP can be implemented at two levels—namely, message-wise and Seb-wise UEP.

\begin{figure}[!t]
    \centering
    \includegraphics[width=0.48\textwidth]{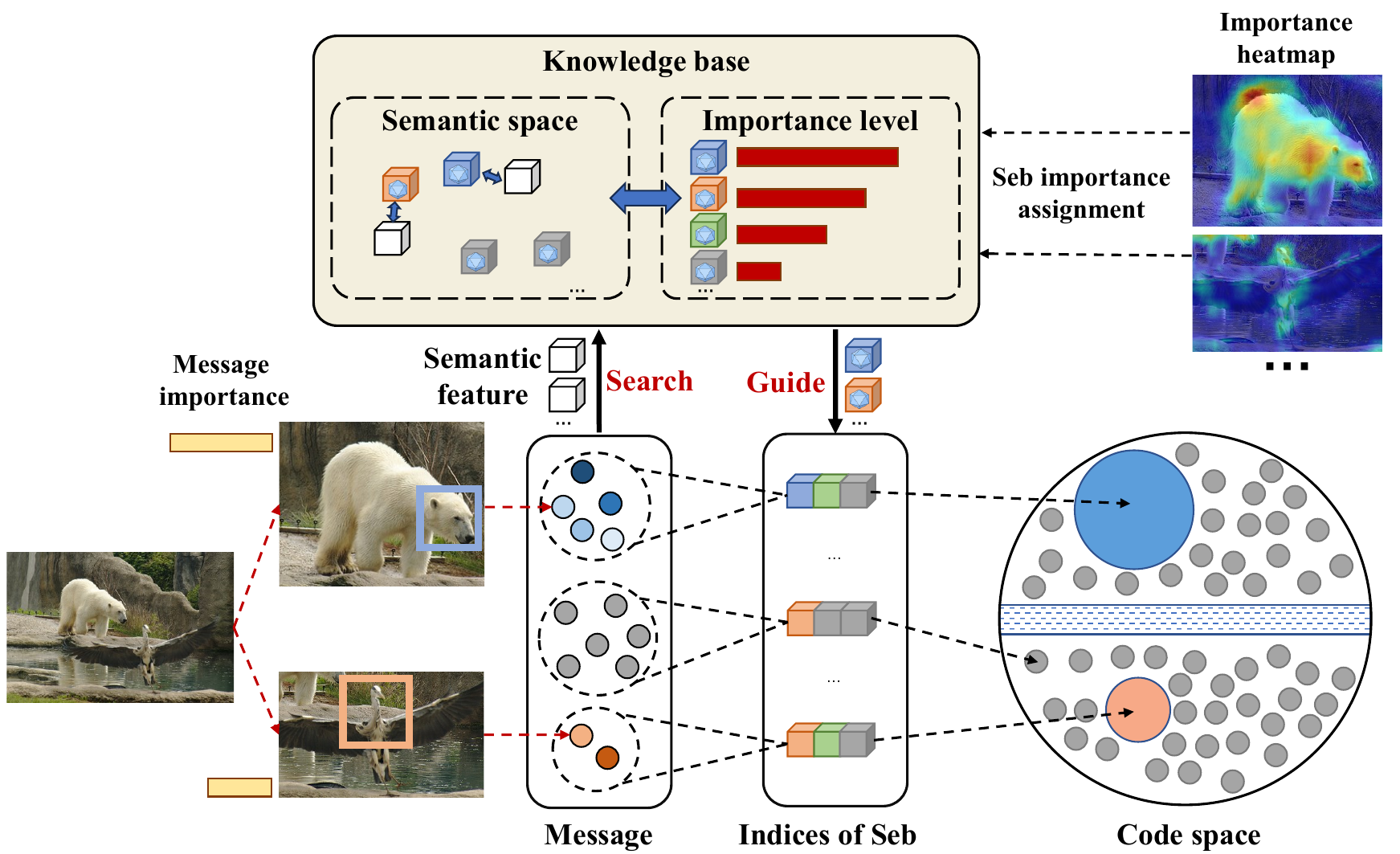}
    \caption{The Seb-wise and message-wise UEP strategies.}
    \label{fig:uep}
\end{figure}

Fig. \ref{fig:uep} illustrates the merits of Seb-wise and message-wise UEP, taking image transmission as an example. Suppose an image contains multiple messages, such as a bear, a bird, and the background in the image. These messages are first mapped to the semantic feature space and represented as symbols by multiple Sebs. Then, these symbols are mapped to the code space, and the mapping process reflects two UEP mechanisms. The message-wise UEP considers the case in which messages have varying importance based on downstream tasks. As shown in the importance heatmap, symbols corresponding to the bear and bird are mapped to larger encoding spheres in the code space, as they are more important than the messages corresponding to the background. The Seb-wise UEP further considers the case in which different Sebs representing a message have distinct levels of importance for a certain downstream task. As illustrated in Fig. \ref{fig:uep}, the blue/orange regions marked in rectangles are distinctive and crucial for classification. Accordingly, the corresponding blue/orange Sebs are assigned a higher level of importance, necessitating prioritized protection. More specifically, the code space is divided into two subspaces, where the Seb-wise UEP mechanism ensures that the Seb corresponding to the bear (i.e., the region marked in the blue rectangle) only appears in the upper half of the code space, while the Seb corresponding to the bird (i.e., the region marked in the orange rectangle) only appears in the lower half of the subspace, ensuring a large distance to avoid confusion between them. Notably, the two levels of UEP can be implemented simultaneously, as they protect semantic information that is crucial for downstream tasks from different perspectives, thereby increasing the robustness of the entire SemCom system.

By employing explicit Sebs and UEP mechanisms, the proposed framework establishes a connection between messages and the knowledge within the KB, which ensures the accurate transmission of critical semantic information, thereby meeting the demands of various intents with greater flexibility. The entire process of KB initialization and communication is summarized in Algorithm \ref{alg1}.

\begin{algorithm}
\caption{KB initialization and communication process in the Seb-based SemCom framework.}
\label{alg1}
\begin{algorithmic}[1]
\Statex \textbf{KB initialization:}
\Statex \textbf{Input: accessible message set $\mathcal{X}$}
\State Initialize the KB with hierarchical Sebs $\langle \mathcal{S},\preccurlyeq \rangle$ according to Eqs. (\ref{eq:clus}) and (\ref{eq:clus_higher}), and attach the set of importance levels of Sebs $\mathcal{I}$ according to Eqs. (\ref{imp}) and (\ref{imp_higher})
\State Synchronize the KB among the UEs
\Statex \textbf{Communication:}
\Statex \textbf{Input: accessible message set $\mathcal{X}$, message to be transmitted at UE $\boldsymbol{x}$, indicator of update request $R=0$}
\State UE estimates accuracy of SR according to Eq. (\ref{eq:acc}) for $\boldsymbol{x}$ and sets update request $R=1$ if SR accuracy $<$ predefined threshold
\State UE performs semantic/channel encoding according to Eqs. (\ref{eq:semen}) and (\ref{chanen}) and sends the message
\If{update request $R=1$}
\State Generate new candidate Sebs $\boldsymbol{s}$ from $\mathcal{X}$ according to Eqs. (\ref{eq:clus}) and (\ref{eq:clus_higher})
\If{$D(\boldsymbol{s}, \mathcal{S})>$ predefined threshold}
\State Integrate new Sebs $\boldsymbol{s}$ into the KB
\EndIf
\State Adjust $\mathcal{I}$ according to Eq. (\ref{eq:adj_imp})
\State Remove obsolete Sebs
\State Synchronize the KB among the UEs
\EndIf
\end{algorithmic}
\end{algorithm}

\section{Protocol of SemComs with explicit Sebs}\label{sec-iii}

\begin{figure}[!t]
    \centering
    \includegraphics[width=0.49\textwidth]{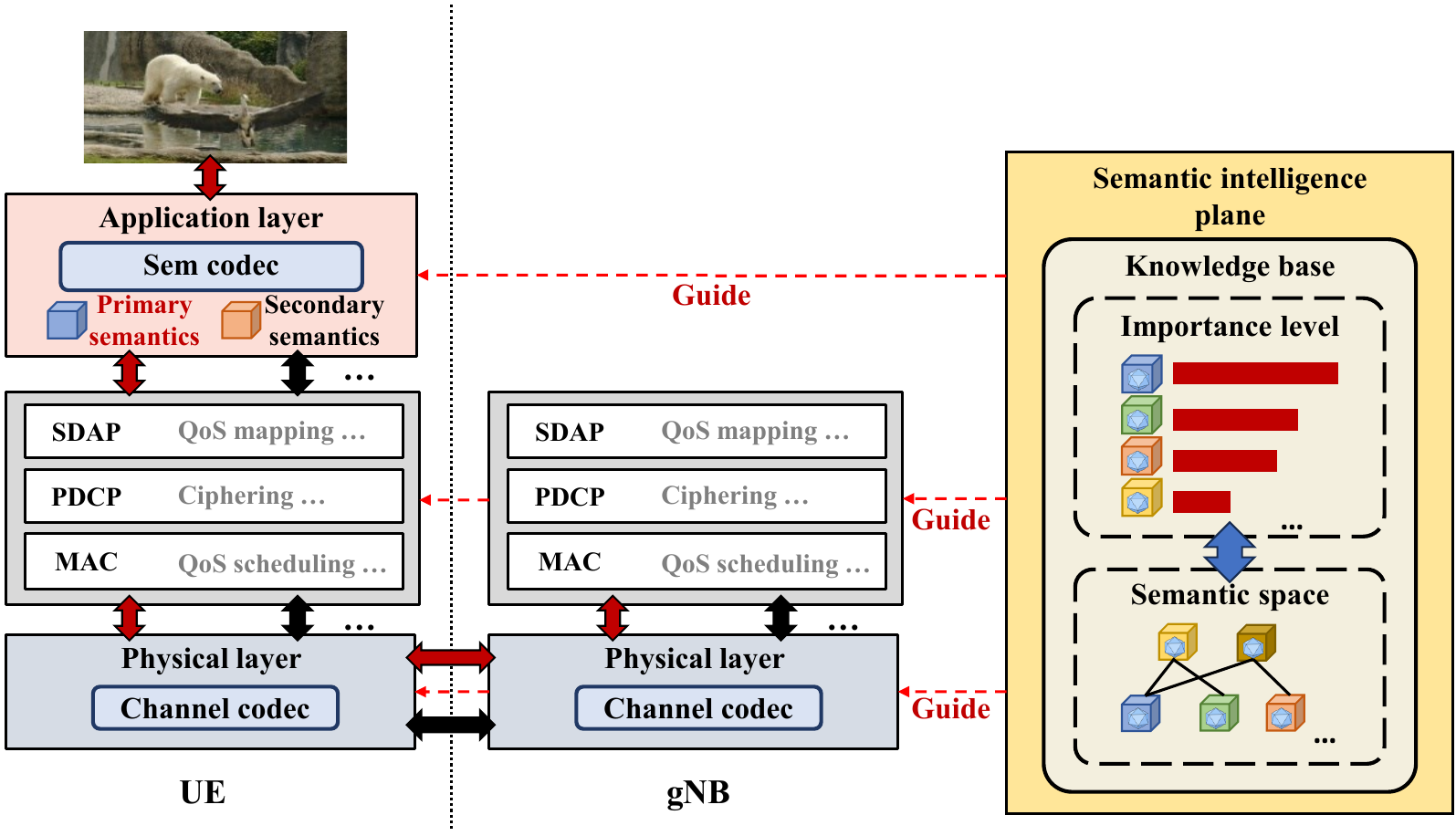}
    \caption{Protocol of SemComs with explicit Sebs. SDAP: service data adaptation protocol; PDCP: packet data convergence protocol; MAC: media access control; QoS: quality-of-service; gNB: next-generation node B.}
    \label{fig:protocol}
\end{figure}

In this section, a protocol stack of the proposed Seb-based SemCom is proposed, which is designed to be highly compatible with the existing 3GPP protocol stack, as shown in Fig. \ref{fig:protocol}. To ensure the system can achieve both intelligence and high efficiency, a semantic intelligence (SI) plane integrating the KB is introduced, where Sebs and their corresponding importance information (as discussed in Section 2) are incorporated into the KB. The introduction of the SI plane establishes interactions between the application and physical layers of the protocol stack, thereby extending the functionality of the 3GPP protocol stacks to process information at the semantic level.

In the uplink process, the message $\boldsymbol{x}$ is first processed by the semantic encoder at the UE’s application layer, where the elements of the output $\boldsymbol{s}^\mathrm{idx}$ are then assigned varying levels of importance. During the semantic encoding process, the SI plane provides Sebs for quantifying raw semantics and synchronizes the importance information with the semantic encoder, enabling the partitioning of $\boldsymbol{s}^\mathrm{idx}$ based on Seb or message importance. Despite being implemented via an NN, the output $\boldsymbol{s}^\mathrm{idx}$ can be represented by bit sequences corresponding to the indices of the Sebs. As a result, the encoded bit sequences can be directly processed by existing protocols such as the service data adaptation protocol (SDAP), packet data convergence protocol (PDCP), and media access control (MAC) for purposes of transmission—including quality-of-service (QoS) mapping, encryption, and scheduling—before the final generation of physical signals. To achieve UEP for semantic flows according to their level of importance, the interaction between the SI plane and the physical layer is taken into consideration. The SI plane provides importance-related information for the upper-layer packets, thereby matching each packet with an appropriate modulation and coding scheme (MCS) and prioritizing the transmission of more critical semantics.

Meanwhile, the next-generation node B (gNB), which shares the same underlying protocol structure as the UE, decodes the received signals and further forwards the decoded packets to the target destinations. The uplink transmission procedure with the proposed Seb-empowered protocol is summarized in Algorithm \ref{alg2} for clarity. Similarly, for downlink scenarios, the SI plane provides Sebs and the corresponding importance-related information for both the gNB and the UE, enabling semantic interpretation of the received signals for the specific intelligent task.

\begin{algorithm}
\caption{Uplink transmission procedure with the proposed SemCom protocol stack.}
\label{alg2}
\begin{algorithmic}[1]
\Statex \textbf{UE Processing:}
\State \textbf{Application Layer:} Encode source message $\boldsymbol{x}$ following Section \ref{sec-ii-sem} with Sebs provided by SI plane, obtaining  according to Eq. (\ref{eq:semen})
\State \textbf{SDAP Layer:} Perform QoS mapping according to importance levels provided by SI plane
\State \textbf{PDCP Layer:} Cipher upper-layer packets
\State \textbf{MAC Layer:} Perform mapping between logical channels to transport channels
\State \textbf{Physical Layer:} Encode upper-layer packets into transmitting data steams using channel encoder proposed in Section \ref{sec-ii-uep} according to Eq. (\ref{chanen}), where different MCS are adopted for packets with different importance levels
\Statex \textbf{gNB Processing:}
\State \textbf{Physical Layer:} Decode received signals following Section \ref{sec-ii-uep}, where packets with different importance levels are correspondingly decoded according to Eq. (\ref{chanen})
\State \textbf{MAC Layer:} Perform mapping between logical channels to transport channels
\State \textbf{PDCP Layer:} Decipher data packets
\State \textbf{SDAP Layer:} Resolve data packets of different QoS flows and prepare the forwarding packet to destination gNB
\end{algorithmic}
\end{algorithm}

\section{Case study: Seb-based SemCom for image transmission}\label{sec-iv}

\begin{figure*}[!t]
    \centering
    \includegraphics[width=0.98\textwidth]{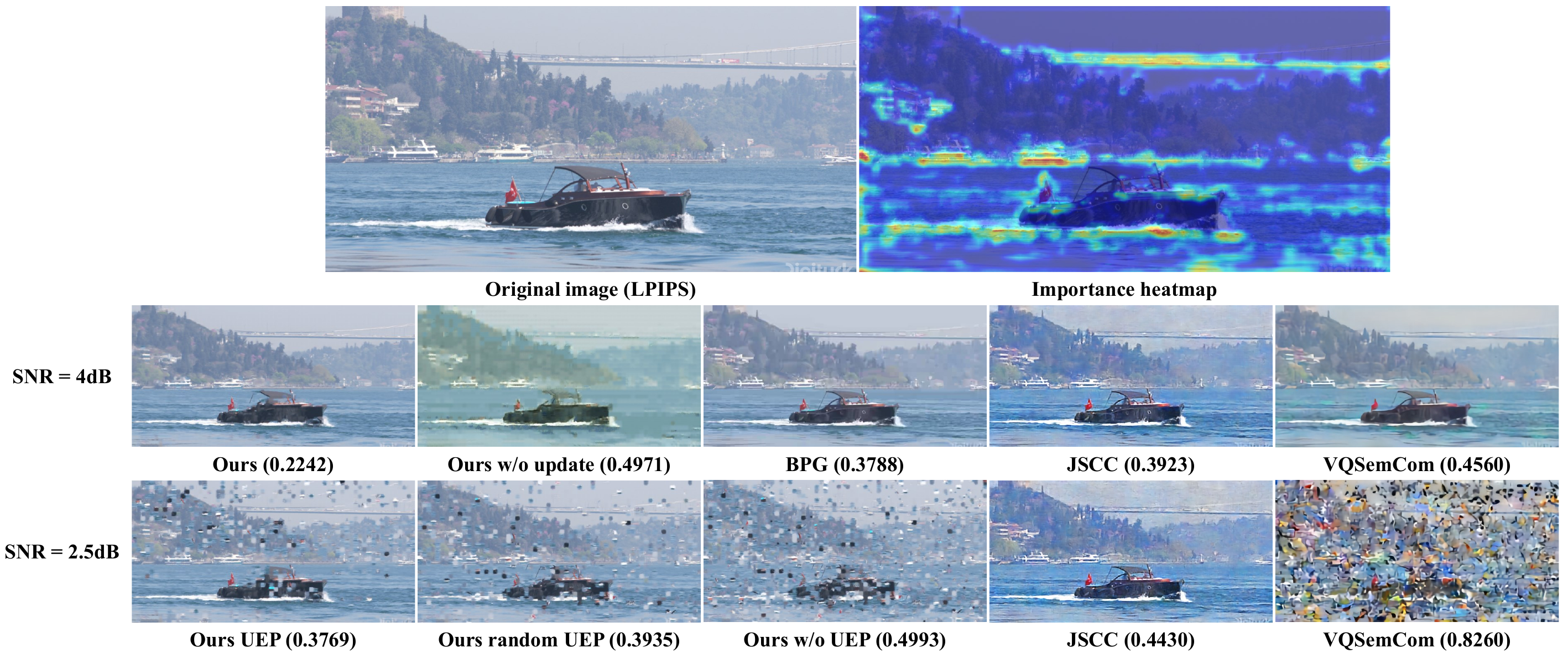}
    \caption{A visual comparison of image reconstruction by different methods. w/o: without.}
    \label{fig:visualization}
\end{figure*}

In this section, a case study of an image-transmission task is carried out to illustrate the performance of the proposed Seb-based SemCom system. The implementation methodology is first introduced; then, the simulation settings and simulation results are presented.

\subsection{Implementation of the SemCom system}
\label{sec-iv-i}

We consider a scenario in which images are transmitted by UEs to the AP, while the AP maintains and synchronizes the KB with the UEs by analyzing the previously received images. The NN of the SemCom system is first trained as an initialization process. Specifically, the semantic extraction module and the task-performing module are implemented by a convolutional neural network (CNN). For semantic extraction, each image is first divided into patches and then processed by four convolution layers with generalized divisive normalization (GDN) \cite{balle2016a} as the activation function to obtain a collection of raw semantics. For task performance, the recovered semantics are inversely processed by four deconvolution layers with inverse GDN (IGDN) \cite{balle2016a} to obtain patch-wise image reconstructions, which are then concatenated \cite{zheng2023seb}. The system is jointly trained on the ImageNet training dataset \cite{imagenet} with 50 000 images being randomly sampled. The learning rate is set to 10\textsuperscript{–4} and 10\textsuperscript{–5} each for one epoch and is then fixed at 10\textsuperscript{–6} until the NN converges.

To realize a hierarchical representation, we consider two different levels of patch size—namely, \{32 $\times$ 32, 16 $\times$ 16\}—to generate coarse and fine-grained Sebs, respectively. For semantic encoding, the top 20\% of the most important regions of the transmitted image are encoded by fine-grained Sebs, and the remaining 80\% of the image regions are encoded by coarse-grained Sebs. For channel encoding, we adopt message-wise UEP. Sebs that represent the top 50\% of the most important regions of the image are transmitted at a lower rate (i.e., using 1/2 rate low-density parity-check (LDPC) and 4-quadrature amplitude modulation (4-QAM)) to ensure higher accuracy. Meanwhile, Sebs representing the remaining 50\% of image regions are transmitted at a higher rate (i.e., using 2/3 rate LDPC and 4-QAM). The test is conducted on additive white Gaussian noise (AWGN) channels. The engineered codec BPG, JPEG2000, and the SemCom methods \cite{bourtsoulatze2019deep} (referred to as “JSCC”) and \cite{fu2023vector} (referred to as “VQSemCom”) are adopted for comparison. To demonstrate the impact of the UEP in our proposed method, all comparison schemes except JSCC directly adopt a high-rate transmission scheme (i.e., using 2/3 rate LDPC and 4-QAM), and the JSCC method is directly trained under an AWGN channel with the target signal-to-noise ratio (SNR; i.e., 4 dB). The channel-bandwidth ratio (CBR) is set equally across all methods by adjusting the source coding rate accordingly for fair comparison. The learned perceptual image patch similarity (LPIPS) metric \cite{zhang2018unreasonable} is primarily adopted to evaluate the quality of the reconstructed images, being defined as the average Euclidean distance between feature maps. The LPIPS effectively reflects human perception, with lower values indicating greater perceptual similarity. The peak signal-to-noise ratio (PSNR) and multi-scale structural similarity (MS-SSIM) \cite{msssim} results are also provided for reference, where higher values indicate better results.

Notably, the proposed Seb-based SemCom system performs well when the testing dataset differs from the training dataset without updating the parameters of the NN. To verify this property, a validation dataset is constructed by sampling images from various distributed datasets. The dataset consists of 30 image subsets, with the first ten subsets randomly sampled from the Cityscapes \cite{cordts2016cityscapes} test set and the middle and last ten subsets randomly sampled from the Bosphorus and YachtRide sequences from the Ultra Video Group (UVG) dataset \cite{mercat2020uvg}, respectively. Each subset contains 30-50 images, each with 2048 $\times$ 1024 pixels. The subsets are transmitted sequentially to mimic cases in which the intent of image transmission changes over time. According to the methodology in Section \ref{sec-ii-kb}, the KB of our method is updated after the transmission of the first, 11th, and 21st subsets; the performance will be validated in Section \ref{sec-iv-ii}.

\subsection{Simulation results}
\label{sec-iv-ii}

A visual comparison of the proposed Seb-based SemCom system with the baselines is shown in Fig. \ref{fig:visualization}, where the first row shows the transmitted image and its corresponding heatmap. The second row shows the visualization and LPIPS results under channel conditions that ensure error-free transmission. The third row shows the corresponding results under poor channel conditions, when transmission errors are inevitable. The CBR of the proposed method, the BPG method, the JSCC method, and the VQSemCom method are set equally.

More specifically, the second row in Fig. \ref{fig:visualization} demonstrates the results when the SNR is 4 dB, allowing all methods to ensure error-free transmissions. The proposed method (referred to as “ours”) utilizes fine-grained Sebs to more precisely represent the visually important foreground areas, such as the ship and the splashing waves, while coarse-grained Sebs ensure efficient representation of the background elements, such as the mountains in the transmitted image. Consequently, the proposed method achieves superior perceptual quality compared with the BPG, JSCC, and VQSemCom methods. The “ours w/o update” scheme refers to the Seb-based framework in the absence of KB updates. Under this condition, distortions in color and detail can be observed due to the mismatch between the Sebs and the current communication intent. It should be noted that we study an extreme case, in which there is a large distinction between the transmission intents, to elucidate the impact of KB synchronization on system performance. In practical applications, intents generally do not undergo such distinct changes, resulting in only a gentle degradation of performance, which will be further illustrated in Section \ref{sec-iv-iii}.

The third row in Fig. \ref{fig:visualization} illustrates the results when SNR is set as 2.5 dB, where transmission errors become inevitable in this case. The BPG method fails to reconstruct the image under this condition and thus is not shown in the figure. In contrast, the SemCom methods demonstrate better robustness against transmission errors. Notably, the JSCC method exhibits the highest robustness, with an LPIPS degradation of 0.05. This robustness is attributed to the joint design of source and channel coding. The VQSemCom method, which employs separate source and channel coding, suffers more severe noise impact, barely revealing the outline of the scene. Our method (referred to as “ours UEP”) effectively mitigates noise impact by emphasizing the protection of the critical image regions highlighted in the heatmap; thus, it maintains an advantage over the JSCC and VQSemCom methods in LPIPS. In contrast, randomly protecting regions (referred to as “ours random UEP”) or transmitting the entire image at a higher rate (referred to as “ours w/o UEP”) results in a more severe noise impact, indicating the effectiveness of the explicit UEP strategies proposed in Section \ref{sec-ii-uep}.

\begin{figure*}[!t]
    \centering
    \subfloat[]{\includegraphics[width=0.33\textwidth]{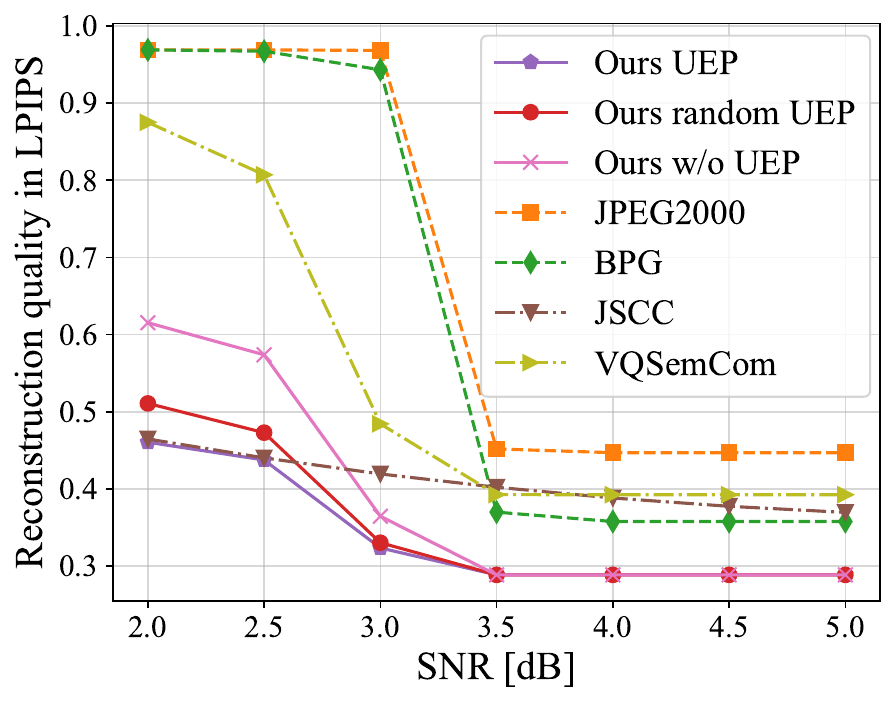}\label{fig:uep_lp}}
    \subfloat[]{\includegraphics[width=0.33\textwidth]{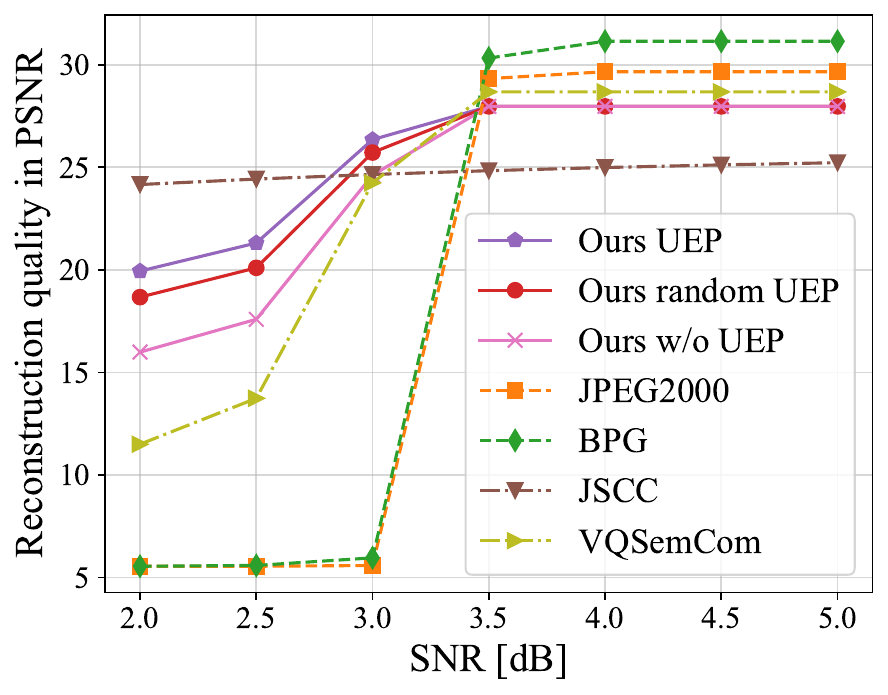}\label{fig:uep_ps}}
    \subfloat[]{\includegraphics[width=0.33\textwidth]{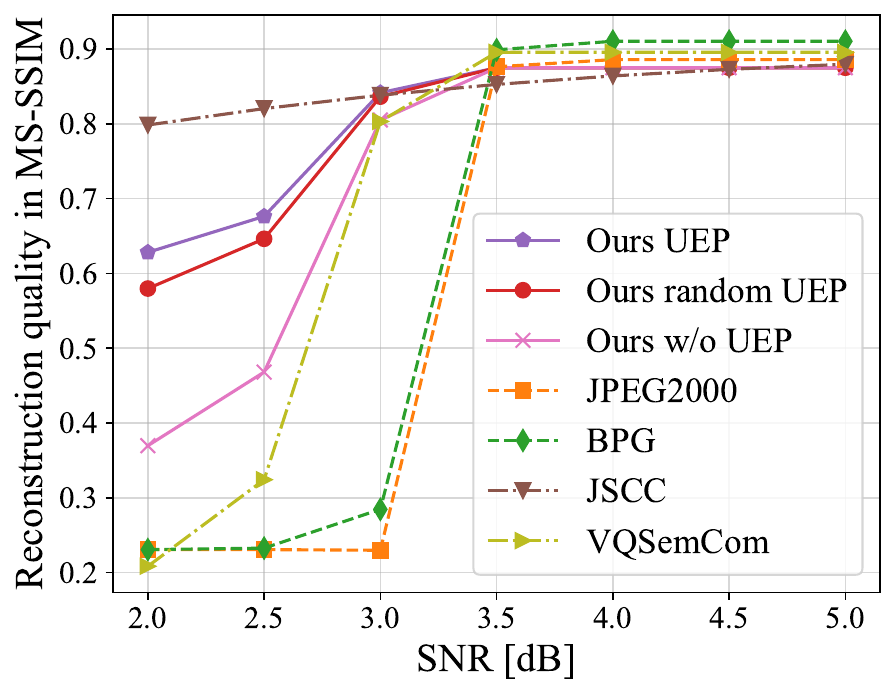}\label{fig:uep_ms}}
    \caption{Overall performance under different SNRs in an AWGN channel. (a) LPIPS under different SNRs. (b) PSNR under different SNRs. (c) MS-SSIM under different SNRs.}
    \label{fig:uep_result}
\end{figure*}

Fig. \ref{fig:uep_result} shows the impact of the SNR on the proposed Seb-based SemCom system. It can be observed that, when the SNR reaches 4 dB and above, all schemes achieve error-free transmission. The proposed method provides more desirable image perceptual quality, with an average improvement of more than 20\% in LPIPS compared with both traditional and SemCom counterparts at the same transmission cost. When the SNR falls below 3.5 dB, the high-rate transmission scheme fails to ensure error-free transmission, leading to a performance decline of differing degrees across all methods. Notably, the proposed Seb-based method with UEP applied (denoted as “ours UEP”) generally outperforms the counterpart that applies UEP randomly to the image (denoted as “ours random UEP”), demonstrating the effectiveness of the importance-driven UEP mechanism. The reason for this effectiveness is that the UEP mechanism enables the Seb-based method to provide additional protection to pivotal parts of the message through the low-rate transmission scheme, further increasing the robustness against channel noise. Moreover, it can be seen that the performance of the BPG and JPEG2000 methods deteriorate more rapidly, showing a “cliff effect” in terms of LPIPS. The VQSemCom method exhibits better robustness compared with traditional methods, as it does not use entropy coding in its encoding process, thereby avoiding further performance degradation brought by error propagation. However, its performance still shows a significant decline. Notably, the Seb-based method, which also uses the same high-rate transmission scheme (denoted as “ours w/o UEP”), despite experiencing severe transmission errors, exhibits better performance than the SemCom baselines, highlighting the advantage of the Seb-based Sem-codec. The JSCC method shows the best robustness against noise, which is attributed to its joint design methodology. However, the upper-bound of its performance under good channel conditions is less competitive. The performance is also evaluated in PSNR and MS-SSIM, which show similar trends as in LPIPS. However, it is interesting to observe that the traditional image coding method BPG generally performs better than all SemCom methods in terms of PSNR and MS-SSIM in a high-SNR regime. This result may occur because the two evaluation metrics focus more on the syntactic aspects of images. In the low-SNR regime, the SemCom systems show better performance, indicating their advantage of robustness under harsh transmitting conditions.

\subsection{Impact of KB updating}
\label{sec-iv-iii}

\begin{figure}[!t]
    \centering
    \includegraphics[width=0.4\textwidth]{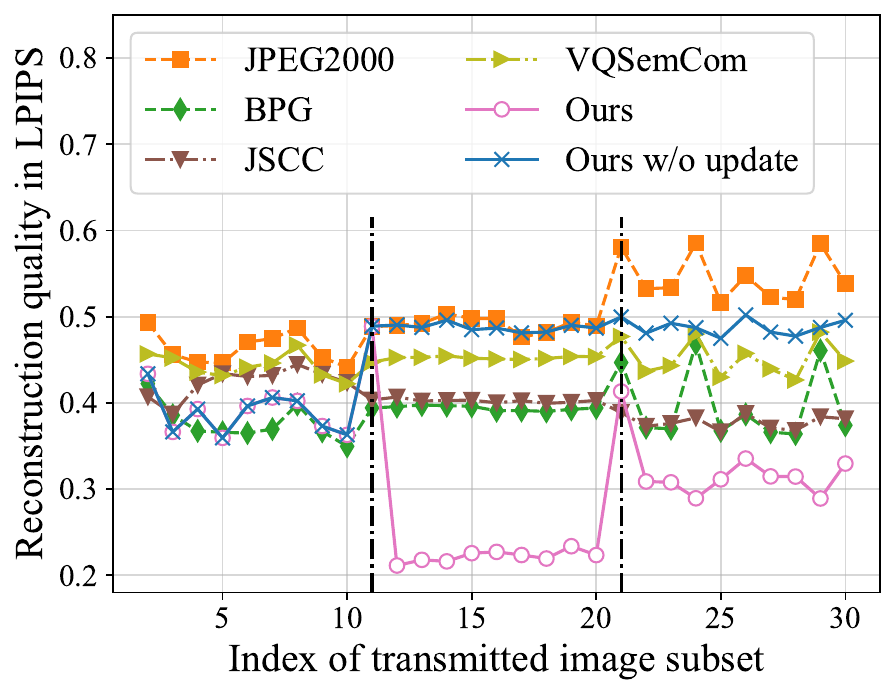}
    \caption{Impact of KB updating, where the communication intent change at the 11th and 21st subsets.}
    \label{fig:kb_up}
\end{figure}

Fig. \ref{fig:kb_up} exhibits the effectiveness of KB updating. The test is conducted under an SNR of 4 dB, ensuring error-free transmissions for the ablation study. The transmission intents change at the 11th and 21st subsets, as depicted in Section \ref{sec-iv-i}. For comparison, the Seb-based system without a KB update is also evaluated, shown as the solid line marked with X’s. It is obvious that the performance degrades once the communication intents change. This occurs because the mismatch between the Sebs in the KB and the changed communication intent results in a poor SR performance in semantic encoding. However, it should be noted that a KB update will be triggered once the SR performance falls below a predefined threshold, as discussed in Section \ref{sec-ii-kb}; the Seb-based SemCom can then quickly evolve without retraining the NN, as shown in Fig. \ref{fig:kb_up}, providing an easier deployment strategy in practical settings.

\subsection{Impact of Seb granularity}
\label{sec-iv-iv}

\begin{figure}[!t]
    \centering
    \includegraphics[width=0.4\textwidth]{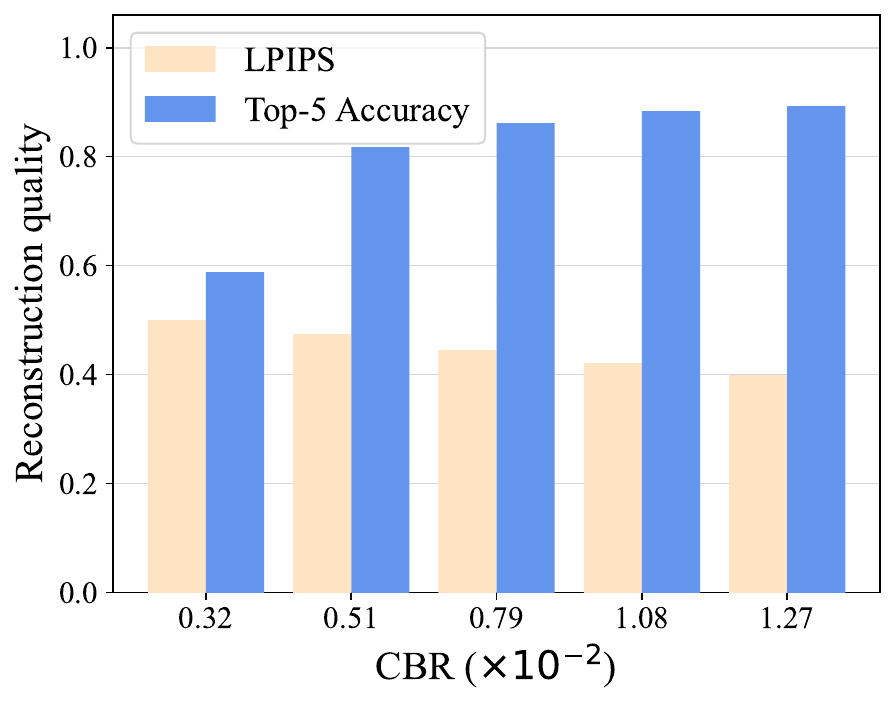}
    \caption{Performance of the Seb-based SemCom with varying ratios of fine-grained Sebs.}
    \label{fig:seb_gran}
\end{figure}

To demonstrate the impact of Seb granularity on the performance of intelligent tasks, an additional experiment is conducted on the Seb-based SemCom system with respect to an image classification task. As shown in Fig. \ref{fig:seb_gran}, different ratios of fine-grained Sebs are adopted to represent the image semantics. The experiment is conducted using the ImageNet \cite{imagenet} validation dataset, which contains 50 image samples per class. Within the dataset, 25 images are randomly selected for Seb generation, while the remaining 25 are used for testing. A ResNet-50 classifier \cite{He_2016_CVPR} is introduced to further perform the image classification tasks based on the reconstructed images. The basic setting of the framework follows Section \ref{sec-iv-i}, except that a uniform 1/2 rate LDPC and 4-QAM are adopted for channel coding to ensure reliable transmission. The ratios of fine-grained Sebs are set as \{0, 0.2, 0.5, 0.8, 1\}, with the corresponding CBR being \{0.32, 0.51, 0.79, 1.08, 1.27\}, represented by the bars from left to right in Fig. \ref{fig:seb_gran}. As the proportion of fine-grained Sebs increases, higher encoding and transmission overheads are incurred.

It can be observed that, as the ratio of fine-grained Sebs increases, the perceptual quality gradually improves, corresponding to the decreases in LPIPS values. This observation is consistent with the analysis in Section \ref{sec-ii-seb}, where fine-grained Sebs provide more precise SR to support higher perceptual quality, at the cost of higher channel bandwidth consumption. The top-5 accuracy with respect to the classification task also increases with the incremental ratio of fine-grained Sebs. However, there is an evident saturation floor with respect to the granularity, which is distinct from the reconstruction task. To be specific, the top-5 accuracy rises from 60\% to 80\% as the proportion of fine-grained Sebs grows from 0\% to 20\%, and the performance is gradually saturated, such that additional increments in the proportion of fine-grained Sebs lead to more moderate gains.

The distinct patterns regarding the impact of granularity for the two different tasks are in line with the analysis in Section \ref{sec-ii-seb}. For the classification task, critical regions within the image contain the key features of objects. The precision of the representation of these critical regions directly determines the accuracy of the classification task. Once the proportion of fine-grained Sebs effectively meets the representation requirements of these regions, the additional increment in fine-grained Sebs no longer contributes to the intelligent task (i.e., classification in this case). As a result, the performance of the intelligent task shows a saturation floor with respect to Seb granularity, where finer-grained Sebs lead to a higher CBR, as shown in Fig. \ref{fig:seb_gran}. In contrast, the reconstruction tasks require detailed semantics across the image; thus, the usage of increasingly finer-grained Sebs leads to consistent improvement in perceptual quality.

Overall, it is obvious that Seb granularity impacts the performance of intelligent tasks, where finer-grained Sebs lead to a better performance. However, finer-grained Sebs also lead to a larger communication overhead—that is, a larger CBR in the experiment. Moreover, the choice of Seb granularity is related to the communication intent—that is, downstream tasks in the system. As a result, the efficiency and overall performance of the Seb-based SemCom can be adaptively controlled by adjusting the Seb granularity in practical use, which is significant for the flexible deployment of the Seb-based SemComs.

\section{Conclusion}\label{sec-v}

In this article, we proposed an explicit Seb-based SemCom architecture for interpretable and flexible semantic-level communications for 6G. First, a mathematical model of Sebs was introduced, followed by descriptions of the designed criteria for the KB, Sem-codec, and channel-codec. A novel Seb-based protocol stack was further proposed, considering the compatibility of current CIT-based networks, with the introduction of an SI plane to incorporate the KB for ubiquitous intelligence and self-evolution for 6G networks. A case study of an image transmission task was conducted, in which the proposed architecture functioned robustly regarding changes in communication intents, obtaining an average improvement of 20\% in the LPIPS compared with the BPG, JPEG2000, JSCC, and VQSemCom baselines. An ablation study showed that the efficiency and overall performance of the proposed Seb-based SemComs can be effectively controlled by adjusting the Seb granularity, demonstrating great flexibility in deployment for practical use.

\bibliographystyle{IEEEtran}
\bibliography{Seb_magazine}

\end{document}